# Heterogeneous dynamic restoration of Ti-15Mo alloy during hot compression

Esmaeil Shahryari [a,*], Maria Cecilia Poletti [a,b], Dalibor Preisler [c], Petr Harcuba [c], Josef Stráský [c], Miloš Janeček [c], Fernando Gustavo Warchomicka [a]

[a] Institute of Materials Science, Joining and Forming, Graz University of Technology, Kopernikusgasse 24/I, 8010, Graz, Austria
[b] Christian Doppler Laboratory for Design of High-Performance Alloys by Thermomechanical Processing, Kopernikusgasse 24, 8010, Graz, Austria
[c] Department of Physics of Materials, Faculty of Mathematics and Physics, Charles University, Ke Karlovu 5, 12116 Prague, Czechia

[*] Corresponding author. Institute of Materials Science, Joining and Forming, Graz University of Technology, Kopernikusgasse 24/I, 8010, Graz, Austria. E-mail address: esmaeil.shahryari@tugraz.at

## Abstract

Near-β titanium alloys have shown low Young's modulus and good strength, making them excellent implant candidates. However, their processing using thermomechanical routes in single phase β region results in heterogeneous microstructures due to high content alloying elements and consequent slow diffusion-controlled processes such as dynamic recovery. This study investigates the deformation behaviour of a Ti-15Mo alloy through hot compression experiments using a Gleeble ® 3800 device in the single β domain at strain rates from 0.01 $s^{-1}$ to 10 $s^{-1}$, reaching final strains of 0.50 and 0.85 followed by immediate water quench. The findings show that the material presents a low strain rate sensitivity. The flow curves show significant strain hardening before reaching a steady-state regime, particularly at high strain rates. The strain hardening exponent calculations support the effect of molybdenum on retarding softening mechanisms such as dynamic recovery. Electron backscatter diffraction (EBSD) measurements of deformed samples revealed that dynamic recovery is the primary restoration mechanism, with continuous and geometric dynamic recrystallisation evidence. Due to the slow restoration process, we observe the subgrain formation for different deformation parameters. Therefore, we introduced an EBSD-based method to quantify dynamic recovery and subgrain size. We concluded that dynamic recovery and dynamic recrystallisation decrease as the strain rate increases with minimal variation at higher strain rates.

## 1 Introduction

Metastable-β titanium alloys have high specific strength, excellent corrosion resistance, good toughness and fatigue strength and can be used in the aerospace, biomedical, and chemical industries [1,2]. Employing biocompatible alloying elements enables these alloys to be used as medical implant materials [3]. Molybdenum (Mo) is a non-toxic and β-stabilising element which reduces the modulus of elasticity [4,5], increases strength and ductility [6], and improves corrosion [7] and wear [8] resistance. Such characteristics make Ti-15Mo suitable for biomedical applications [9–12].

After designing the composition, thermomechanical processing (TMP) is the subsequent step in developing parts with a near-net shape. TMP also enhances the mechanical properties by controlling the microstructure [13,14]. The microstructure evolves due to different deformation mechanisms that depend mainly on the alloy's stacking fault energy (SFE). In high SFE alloys, such as most of the β titanium alloys [15,16], aluminium alloys [17–19], and α-iron [20], dislocations form and, despite partial annihilation, rearrange into subgrain boundaries by the mechanism of dynamic recovery (DRV) [21]. Due to the intense recovery, the dislocation density is insufficient to bulge high-angle grain boundaries that grow to form new grains by discontinuous dynamic recrystallisation (dDRX). In high SFE materials, DRV is accompanied by continuous dynamic recrystallisation (cDRX) and geometric dynamic recrystallisation (gDRX). The cDRX involves the progressive and continuous development of misorientation within low-angle grain boundaries (LAGBs) through lattice rotation, transforming them into high-angle grain boundaries (HAGBs). If grains are geometrically stretched, compressed or squeezed after large deformations, gDRX is responsible for forming grains with sizes comparable to those of the subgrains. The gDRX mechanism involves the serration of HAGBs, decreasing of grain thickness, and impingement of serrated HAGBs [21,22].

Metastable titanium alloys, during hot deformation in the β-phase region, are known to undergo DRV at early-stage strain levels, followed by cDRX and gDRX at large strains. Warchomicka et al. [23] observed DRV and cDRX as the predominant mechanisms during the hot compression of Ti55531. They also reported the local formation of small new grains by gDRX at large strains. Poletti et al. [24] investigated the deformation behaviour of the β phase in Ti64 and Ti55531 alloys. Both alloys' deformation mechanisms are DRV followed by cDRX at high strain rates and solely DRV at low and moderate strain rates, with no evidence of dDRX. Similarly, DRV followed by cDRX was reported by Buzolin et al. during the hot compression of a Ti5553 alloy [25] and a Ti-17 alloy [26]. Matsumoto et al. found that DRV dominates restoration in deformed Ti-5553 [16] and Ti-6246 [27] alloy with some cDRX near prior β-grain boundaries. Other studies have also observed the occurrence of DRV and cDRX during hot compression of titanium alloys [28–34]. However, dDRX, frequently reported in low and medium SFE alloys, is also proposed to coexist with cDRX during the hot compression of titanium alloys [35–39]. The authors concluded that new dDRX grains are formed at triple junctions, particularly at low strain rates. The occurrence of dDRX could be ascribed to factors such as a slow post-deformation cooling rate, which allows for post-dDRX, and the presence of alloying elements that reduce SFE.

Mo is the most effective β-stabilizer, reducing the SFE of titanium [40,41]. Jae-Gwan et al. [42] reported that increasing the Mo content in Ti-xMo-4Fe alloys decelerates the diffusion rate and retards the dynamic recrystallisation during hot compression. However, their study did not distinguish between continuous or discontinuous DRX. Wang et al. [43] discussed that dDRX and cDRX coexist in hot compression deformation in near-α titanium alloy. The observed dDRX might be linked to the relatively low SFE of near-α titanium alloy.

Most studies on Ti-15Mo alloy have focused either on its mechanical properties [14,44], or phase transformation [45–49], resulting in a lack of information about the thermomechanical behaviour of this alloy. Hot compression of Ti-17Mo in the single β region presented by Ebied et al. [50] reported the formation of large grains by dDRX at high temperature and low strain rate and fine grains by cDRX at high strain rate. In addition, they noted that DRV was more active than DRX in restoring the deformed microstructure. The observation of dDRX by the authors at higher temperatures and low strain rates is somewhat surprising since the increased dislocation mobility at higher temperatures and sufficient time allowed at low strain rates favour DRV, which can suppress the occurrence of dDRX.

In this work, we investigate the flow behaviour and the microstructural evolution of a Ti-15Mo alloy during hot compression in the β region at different temperatures and strain rates, focusing on the effect of the presence of Mo atoms in single phase material on the restoration mechanisms. Since we expect a slow DRV rate compared to other alloys, we hypothesise that the DRV rate does not rapidly saturate, as occurs in most of the high SFE materials [17]. Therefore, we developed a method using electron backscatter diffraction (EBSD) measurements to quantify the fractions of dynamically recovered and recrystallised microstructures and the evolution of the subgrain size with respect to the strain rate.

## 2 Methodology

### 2.1 Materials

We used a binary Ti-Mo alloy with 15 wt.% Mo in this study. This alloy was produced by ERCATA GmbH (Germany) by electron beam melting and contains approximately 0.02 wt.% of oxygen and 0.003 wt.% of nitrogen. The material was cast and cold-swaged into circular rods of 11 mm in diameter. The bars were solution-treated at 900°C for 30 minutes in a vacuum chamber and immediately water-quenched to achieve a β-annealed condition. Using differential scanning calorimetry (DSC) measurements, we determined the β-transus temperature of the investigated Ti-15Mo at 735 °C.

### 2.2 Hot compression experiments

Cylindrical specimens with a diameter of 10 mm and a height of 15 mm were machined from the annealed rods. The specimens underwent compression at elevated temperatures using a Gleeble® 3800 thermomechanical processing simulator. Three K-type thermocouples welded at the surface of the specimen gave information about the temperature. The thermocouple welded at the middle of the specimen gauge controlled the temperature during the experiment. The two other thermocouples located at 1 mm from each edge of the specimen measured the temperature gradient along the sample in the compression axis. A combination of graphite foil, tantalum foil and nickel-based paste was placed at the specimen interfaces with anvils to minimise friction and thermal gradient. Before the compression test, the β-annealed specimens were heated up to the deformation temperature at 5 K/s, held for 5 min to ensure the temperature stabilisation before the deformation and then isothermally compressed at a constant strain rate. The deformed samples were in-situ water-quenched after hot deformation to room temperature to preserve the deformed microstructure for further microscopic analyses. The whole compression test process was performed under a protective atmosphere of argon to prevent oxidation of the samples.

Uniaxial hot compression experiments were conducted on Ti-15Mo specimens in the single β-phase region at temperatures of 810 °C, 860 °C and 910 °C and strain rates of 0.01 $s^{-1}$, 0.1 $s^{-1}$, 1 $s^{-1}$ and 10 $s^{-1}$ and up to the two final true strains of 0.50 and 0.85. Additional compression tests were also performed on Ti-15Mo at 930°C and 0.1 $s^{-1}$, solely to compare the strain hardening of this alloy with that of Ti-5Al-2Sn-2Zr-4Cr-4Mo (Ti-17) investigated by Buzolin et al. [26]. The Mo equivalent of Ti-17 was 5.07%, determined using the equation from Cotton et al. [2] for the given chemical composition [26]. A heat treatment experiment was performed at 910 °C for 5 min, and immediately water quenched to examine the microstructure before the deformation.

### 2.2.1 Flow curve correction

A smoothing procedure removed signal noise in the stress-strain data recorded by the Gleeble software in the Origin software v. 2021b. Flow stresses were then corrected for adiabatic heating as described elsewhere [51].

### 2.2.2 Strain-hardening exponent

The strain hardening exponent ($n$) at a constant strain rate and deformation temperature was calculated from the slope of the stress-strain curves using Origin software v. 2021b according to Hollomon's equation (Eq. (1)) [52,53].

$$\sigma = K\varepsilon^n \;\rightarrow\; n = \varepsilon/\sigma \cdot \partial\sigma/\partial\varepsilon|_{\dot{\varepsilon},T} \quad (1)$$

where $K$ is the strength factor, $\sigma$ is the true stress, $\varepsilon$ is the true plastic strain, $\dot{\varepsilon}$ is the strain rate, and $T$ is the temperature. The strain hardening exponent was determined for all deformation conditions.

### 2.2.3 Strain rate sensitivity

The strain rate sensitivity index ($m$) measures the change in flow stress with the variation in the strain rate, which can be expressed as a function of natural logarithmic stress and strain rate values, as shown in Eq. (2) [21].

$$\sigma = K\dot{\varepsilon}^m \;\rightarrow\; m = \partial(\log\sigma)/\partial(\log\dot{\varepsilon})|_{\varepsilon,T} \quad (2)$$

The natural logarithm of true stress was plotted against the natural logarithm of strain rate at a fixed temperature and strain. A linear fit was applied, where the slope of the fitting line denotes the strain rate sensitivity.

## 2.3 Metallography

Undeformed and deformed samples were sectioned in half along the axial axis. The pieces were hot mounted on PolyFast resin and ground using SiC papers (#320 ~ #1200) and diamond suspension in grain size of 9 µm and then polished with a mixture of colloidal silica suspension (OP-S) and hydrogen peroxide ($H_2O_2$) in a ratio of 5:1.

EBSD measurements were conducted to characterise the microstructure in the central region of the deformed samples. A field emission gun scanning electron microscope (FEG-SEM) TESCAN Mira3 was used with an EBSD Hikari camera (EDAX-Ametek) and APEX V2.5.1001 software. EBSD measurements covered an area of 300 µm × 300 µm with a step size of 0.25 µm, using an acceleration voltage of 25 kV, beam intensity of 17 and a working distance between 19 mm and 22 mm. The EBSD data were treated using the OIM Analysis software v.8.6. In the EBSD images presented throughout this study, black lines indicate HAGBs with a misorientation exceeding 12°. In contrast, white lines indicate LAGBs with misorientation within the range of 2° to 12°.

# 3 Results

## 3.1 Microstructure before hot deformation

The Inverse Pole Figure (IPF) maps in Figure 1 show almost fully recrystallised microstructures featuring equiaxed grains. The average grain size of the annealed sample is 88±32 µm, increasing to 109±42 µm after the sample is heat-treated in Gleeble for five minutes at 910 °C. This represents approximately 24 % grain growth compared to the solely annealed condition. The grain growth before deformation at lower temperatures (810 °C and 860 °C) was also examined, and it is lower than at 910 °C.

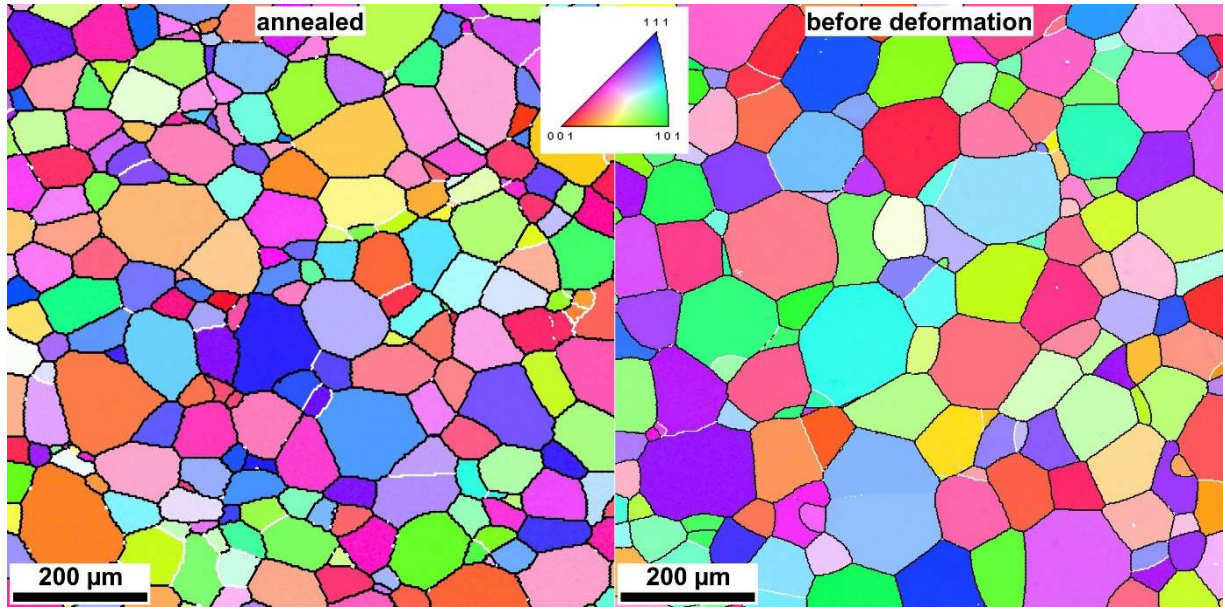

**Figure 1. IPF maps of an annealed sample and a sample before deformation at 910 °C.**

## 3.2 Flow stress

Figure 2(a-c) displays the smoothed and adiabatic-heat corrected flow curves of the investigated Ti-15Mo alloy obtained through compression at 810 °C, 860 °C, and 910 °C and strain rates between 0.01 $s^{-1}$ and 10 $s^{-1}$. The flow curves were obtained up to two final strains of 0.50 and 0.85, confirming a good reproducibility of the experiments with a maximum relative error of about 9.8 % corresponding to the deformations at 860 °C and 10 $s^{-1}$. Higher temperature and lower strain rate both reduce flow stress. Higher temperature provides the energy for faster dislocation motion, while lower strain rate gives time for dislocation rearrangement and annihilation. Figure 2d exhibits an almost steady state regime after a slight discontinuous yielding in the flow curves at 0.01 $s^{-1}$, while at higher strain rates, they are characterised by a broad work hardening up to the strain of 0.2 ~ 0.3. The broad work hardening reaches a steady state at 0.1 $s^{-1}$ but increases slightly at higher strain rates.

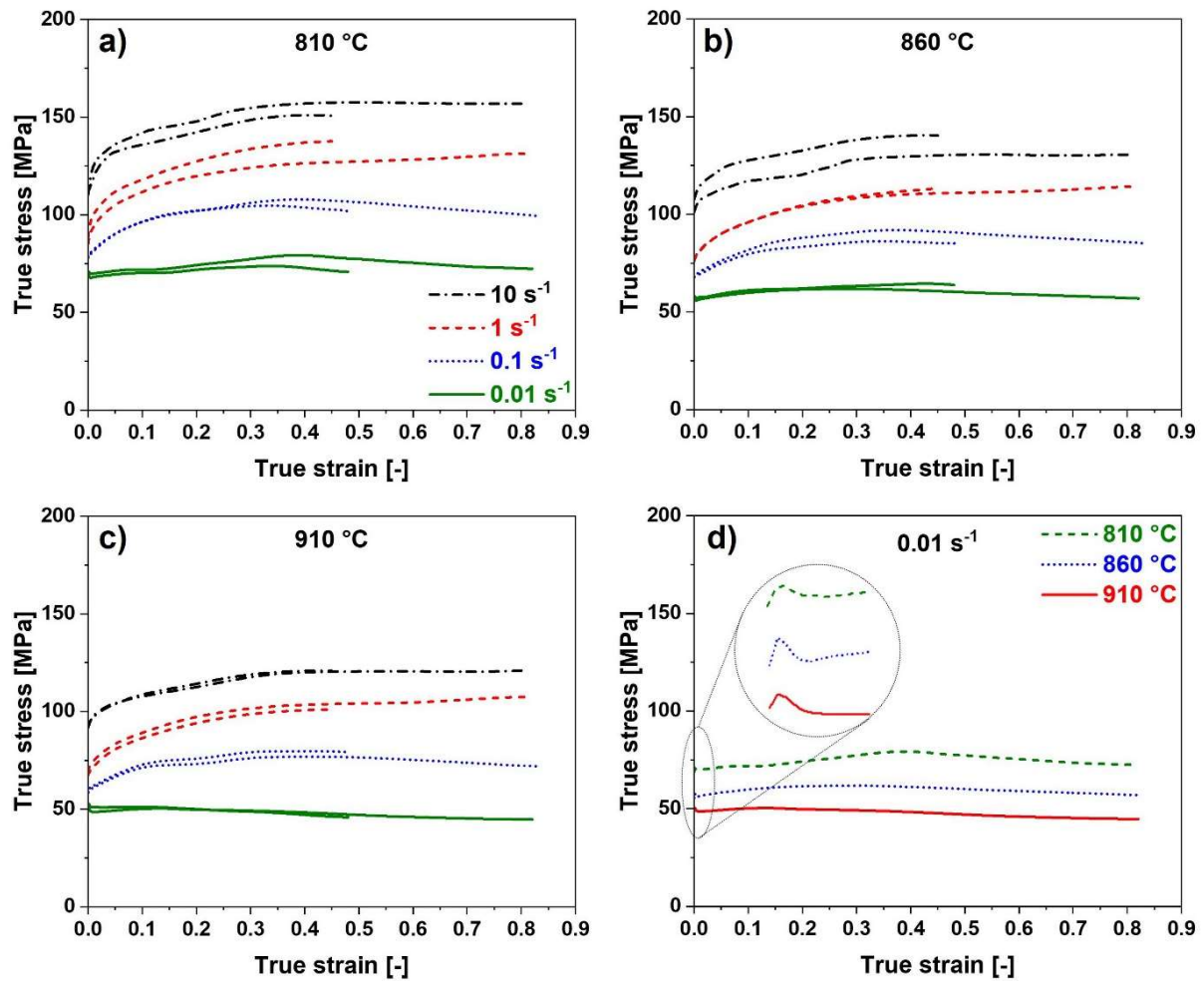

**Figure 2. Flow curves of the studied Ti-15Mo alloy at various strain rates and temperatures.**

### 3.2.1 Strain-hardening exponent

Figure 3(a-c) illustrates the evolution of the strain hardening exponent with the strain for the studied Ti-15Mo alloy under compression across all deformation conditions. Initially, the exponent increases sharply due to the rapid multiplication of dislocations during the early stages of deformation, reaching a peak value. Then, it decreases as the contribution of the softening mechanism becomes more prominent and eventually tends to the zero value at peak strains when the role of the softening mechanisms balances that of the work hardening. Notably, deformation at 0.01 s$^{-1}$ exhibits less strain hardening than at higher rates because more time at lower strain rates allows enhanced softening mechanisms. Figure 3d also illustrates that the strain hardening exponent reaches zero at lower peak strains as the strain rate decreases, indicating the enhanced DRV under these conditions, as shown by Momeni et al. [54].

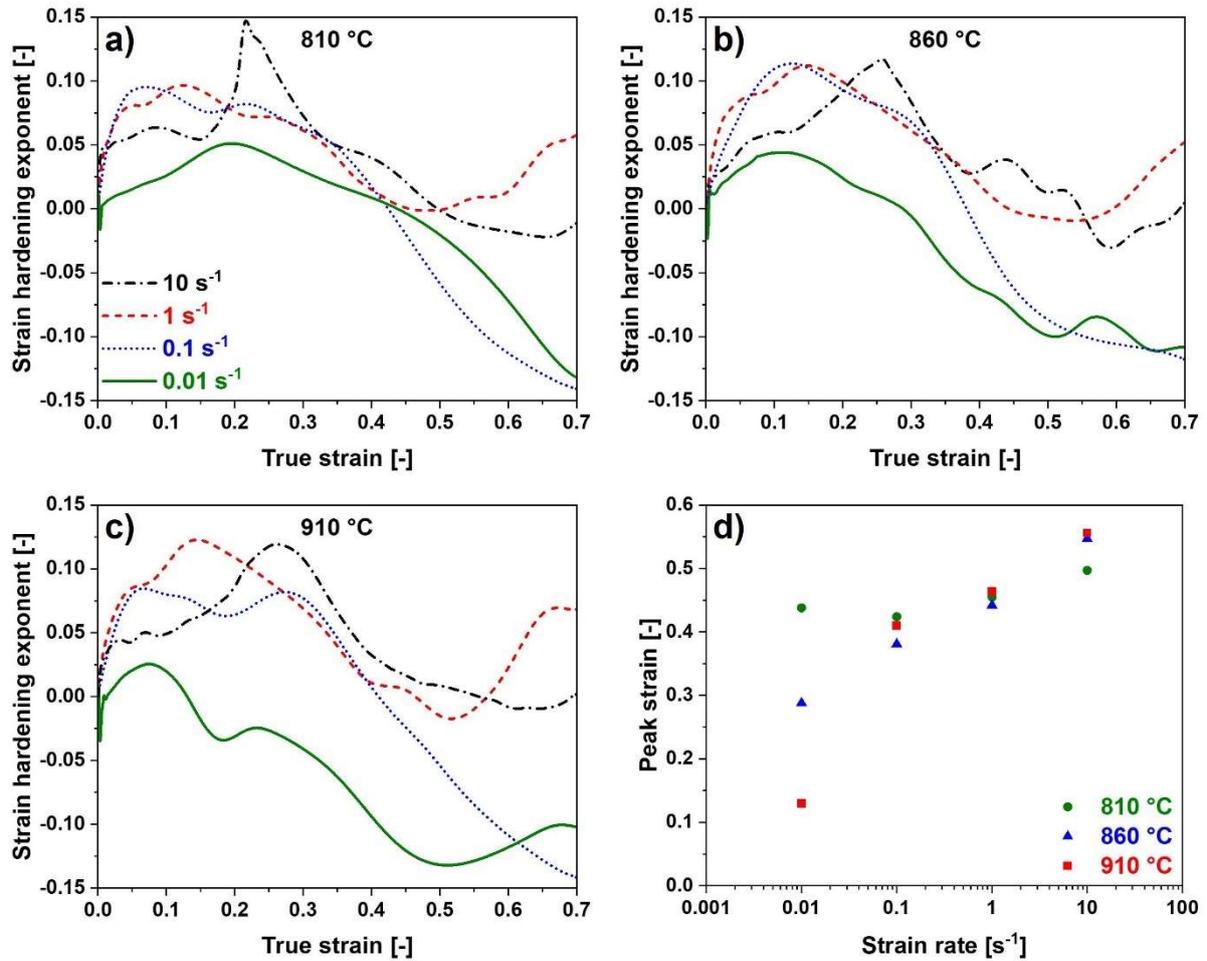

**Figure 3. a-c) Strain hardening exponent, d) Peak strain, at various strain rates and temperatures.**

### 3.2.2 Strain rate sensitivity

Figure 4a shows the natural logarithm values of the true stresses and strain rates at a strain 0.50 and three test temperatures. The slope of the fitted lines represents the strain rate sensitivities. The strain rate sensitivity at 0.50 of strain changes slightly from 0.10 at 810 °C to 0.13 at 910°C. Figure 4b shows the strain rate sensitivities versus the temperature at different strain values. The strain rate sensitivity of Ti-15Mo alloy is relatively low compared to values for typical near-β titanium alloys [13,55,56]. Ebied et al. [50] determined an average stress exponent of 8.04 for hot compression of Ti-17Mo, corresponding to an average strain rate sensitivity of 1/8.04 ≈ 0.124, which agrees with our results.

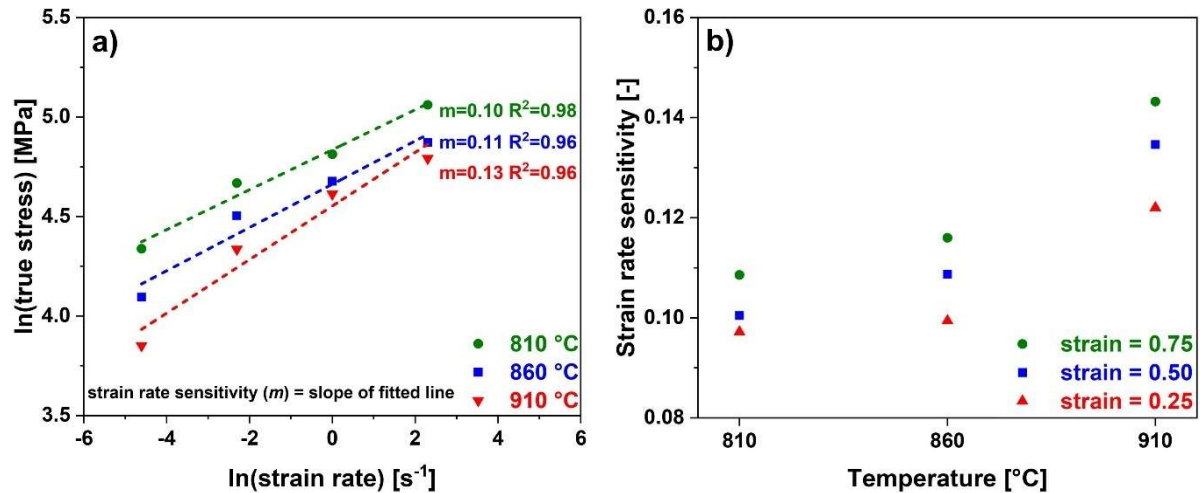

**Figure 4. a) Strain rate sensitivity at a strain of 0.50, b) Variation of the strain rate sensitivity with temperature.**

### 3.3 Microstructure after deformation

Figure 5 (a-f) shows the IPF maps depicting the microstructure of the studied Ti-15Mo alloy compressed up to a final strain of 0.85 at various temperatures and strain rates. The rest of the IPF images can be found in the supplementary data (Figures S1-S3). Different deformation characteristics and restoration mechanisms can be inferred from Figure 5. The prior grains were elongated significantly perpendicular to the compression load direction at all deformation conditions. The subgrains form within elongated grains due to DRV. The degree of deformation within grains is heterogeneous due to the different initial crystallographic orientations of the grains. The HAGBs serrate with a wavelength close to the subgrain size. Such serrations, attributed to local migration of HAGBs, indicate that gDRX can occur where a flattened grain's thickness reaches the subgrain size. Through gDRX, a pre-existing grain is pinched by serrated HAGBs, forming new recrystallised grains with the size of subgrains. The recrystallised grains formed by gDRX have orientations almost identical to those of the parent grains. Examples of new grains formed by gDRX are indicated by arrows in Figure 5a. This figure also highlights visual evidence of the early stage of cDRX, particularly in regions where some LAGBs have been transformed into HAGBs through lattice rotation, forming subgrains with partially developed HAGBs. Figure 5f also shows more new grains formed by cDRX. New grains formed by cDRX show different orientations due to lattice rotation. Our investigation has found no evidence of dDRX. Figure 5 (a-d) shows the strain rate's effect on the microstructure's evolution. It compares the deformed microstructures at 860 °C at strain rates ranging from 0.01 s$^{-1}$ to 10 s$^{-1}$. The area covered by subgrains is larger for deformations at 0.01 s$^{-1}$ (Figure 5a) than in samples deformed at strain rates of 0.1 s$^{-1}$ and 1 s$^{-1}$ (Figure 5 (b, c)). As the strain rate increases to 10 s$^{-1}$, some prior grains exhibit a lean presence of subgrains (Figure 5d). Figure 5 (e, f) compares the effect of temperature on the microstructure evolution of samples deformed at 0.1 s$^{-1}$. This comparison demonstrates a notable increase in the size and volume fraction of recrystallised grains with a temperature rise of 100 °C.

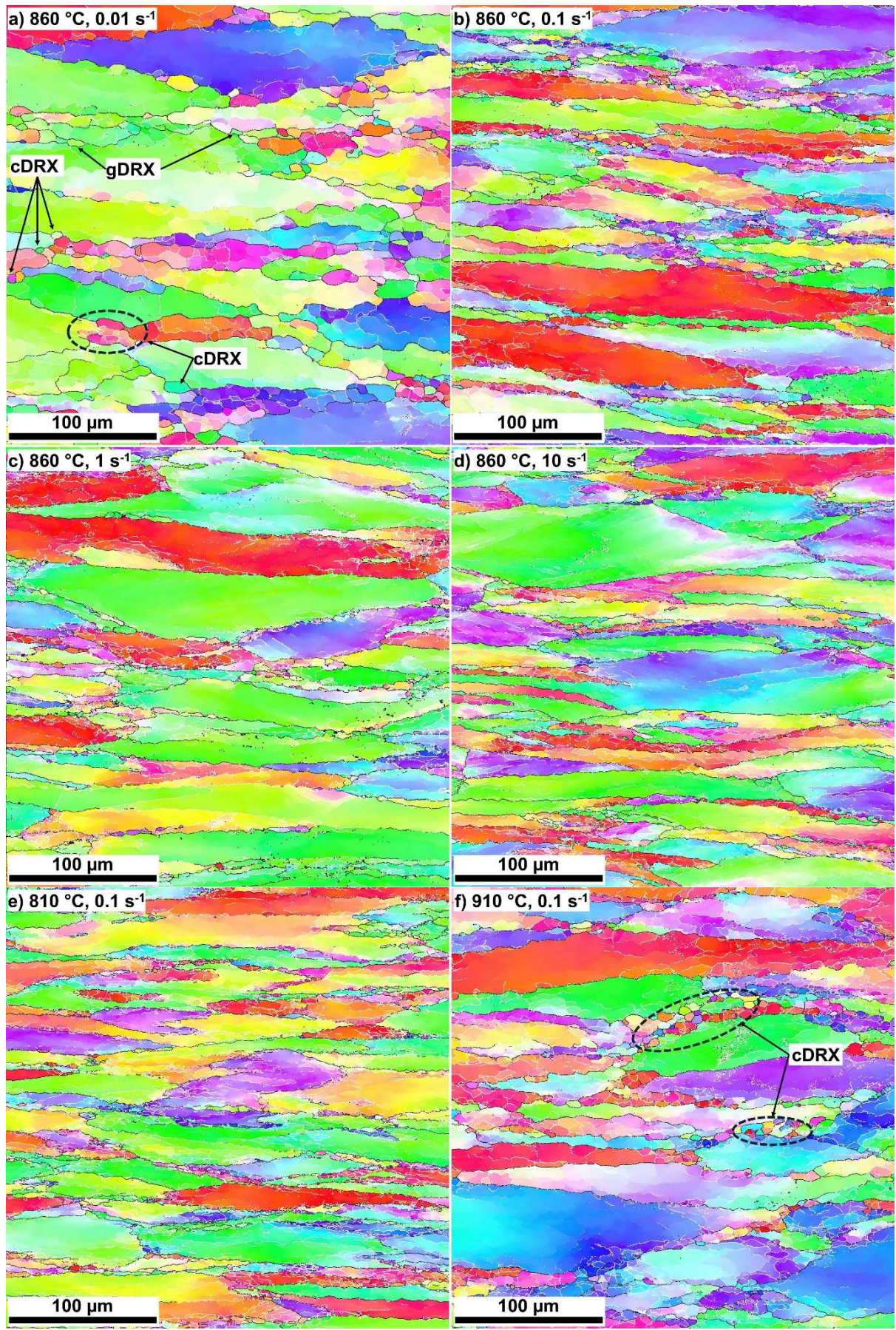

**Figure 5. IPF maps of samples deformed at different strain rates and temperatures (final strain = 0.85). Compression axis is vertical.**

## 4 Discussion

### 4.1 Occurrence of dynamic recovery and dynamic recrystallisation

The steady-state regime of flow stress, mainly observed at lower strain rates, indicates a dynamic restoration mechanism in equilibrium with the observed work hardening. EBSD-IPF maps (Figure 5) of the deformed samples also reveal that the dominant deformation mechanism is DRV, characterised by the formation of subgrains, followed by the early stages of cDRX and gDRX occurring concurrently. More subgrains are formed during deformation at low strain rates due to sufficient time available for the rearrangement of dislocations. Grain boundary and dislocation mobilities are higher during deformation at higher temperatures. Consequently, subgrains form within the grains faster due to the accelerated DRV. At higher strain rates, the rapid increase in dislocation density attributed to the short time available for DRV limits the formation of subgrains and misorientation, which spread only close to the prior HAGBs. Due to adequate time for the local motion of HAGBs at lower strain rates, the grain boundary serration exhibits a higher amplitude than that found at the same deformation temperature and a higher strain rate. The more pronounced serration results in more grain impingements, forming more new grains via gDRX.

Additionally, the extent of cDRX increases at lower strain rates by progressive rotation of subgrains. However, both recrystallisation mechanisms only occur at high levels of strain. Comparable observations have been documented in the literature [25,26,55,57]. Furthermore, Ebied et al. [50] reported the occurrence of dDRX beside cDRX during hot compression of Ti-17Mo in the β single phase field. In contrast, we observed a similar microstructure in Ti-15Mo but without evidence of dDRX.

#### 4.1.1 Quantification of dynamic recovery and dynamic recrystallisation

In this work, we observed a highly heterogeneous microstructure after deformation. Therefore, we introduce a method for quantifying dynamic recovery and dynamic recrystallisation using EBSD measurements. This method defines a subgrain as a region surrounded by boundaries with a misorientation angle exceeding 0.6°. This value was used in the OIM software to determine the grain tolerance angle to separate subgrains. Furthermore, we excluded grains smaller than 1.0 µm to minimise noise.

The Grain Orientation Spread (GOS) was used to determine the total fraction of recovered and recrystallised regions by selecting an appropriate absolute range of GOS for each deformation condition, typically from 1° to 2°. This very low value refers to well-recovered subgrains and new grains formed via cDRX and gDRX. Therefore, the area determined by low GOS is referred to as the restored area, representing the sum of recovered and recrystallised fractions. We visually examined the recovered subgrains and new recrystallised grains for each deformation condition with the IPF image. We varied the GOS limit by 0.05 degrees to ensure that only dynamically recovered and dynamic recrystallised regions were captured. This somewhat manual process improves the accuracy of the determined fraction of the restored area and limits the average error to less than 2%. For example, Figure 6a shows an Image Quality (IQ) map overlapped with GOS, highlighting the restored regions in brown for the sample deformed at 860 °C and 0.1 s$^{-1}$ up to the final strain of 0.85.

Conversely, the fraction of only dynamic recrystallised area is determined by setting the grain tolerance angle to a rather high value of 12°. This higher tolerance angle identifies regions with significant misorientation angles, which is characteristic of recrystallised grains. The upper limit of GOS of these grains is found in a low range from 1.20° to 2.26°, confirming that the grains are indeed recrystallised (blue regions in Figure 6b). We subtracted the recrystallised fraction from the restored fraction to determine the recovered fraction. The GOS maps showing the restored and recrystallised fractions for all the studied parameters are available in the supplementary data (Figures S1-S3).

Some other studies proposed using fixed GOS values, such as 6° [58] and 2° [32,42,59] to distinguish recovered and recrystallised areas, respectively. It should be noted that although the GOS angles used in this study are in the range of 1° to 2°, they are not directly comparable to the values mentioned above in the literature, as we have used different grain tolerance angles compared to the default 5° which affects the measured GOS within the grains/subgrains.

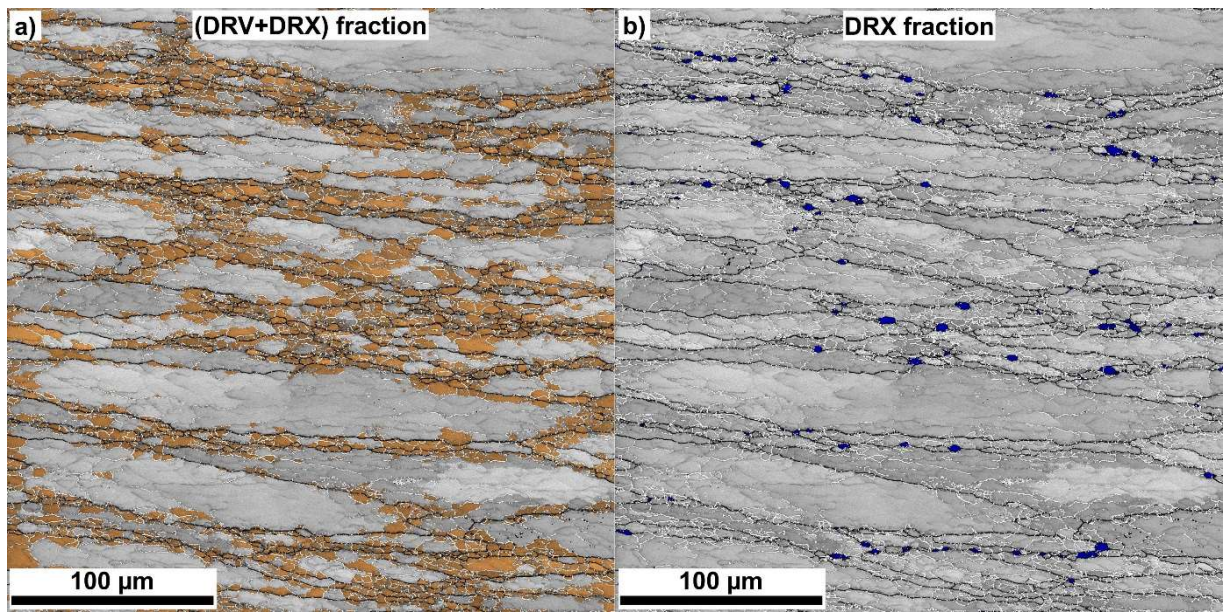

**Figure 6.** Maps of IQ+GOS showing a) (DRV+DRX) area and b) DRX area corresponding to compression at 860 °C and 0.1 s$^{-1}$ up to a final strain of 0.85. Compression axis is vertical.

Figure 7a shows that the area fraction of DRV, expressed as a percentage, is plotted against strain rates at three different temperatures. This plot corresponds to the deformations up to the final strain of 0.85. As expected, the recovered area fraction depends on deformation temperature and strain rate. Higher temperatures lead to a larger recovered area at a constant strain rate. Conversely, a higher strain rate results in a lower recovered area at a given temperature. However, there is minimal change in the recovered area when the strain rate increases from 1 s$^{-1}$ to 10 s$^{-1}$. The maximum recovered area was 37.2% for deformation at the highest temperature (910 °C) and the lowest strain rate (0.01 s$^{-1}$), meaning that DRV is not completed for the achieved strains even under these conditions. Figure 7b shows the variation of the DRX fraction, representing the combined contribution of both cDRX and gDRX for different deformation conditions. The variation is negligible despite a decrease in DRX fraction with both lower temperature and higher strain rate. The DRX fraction is low, with a maximum value of only 2% at 910 °C and 0.01 s$^{-1}$, indicating that DRV is the primary restoration mechanism.

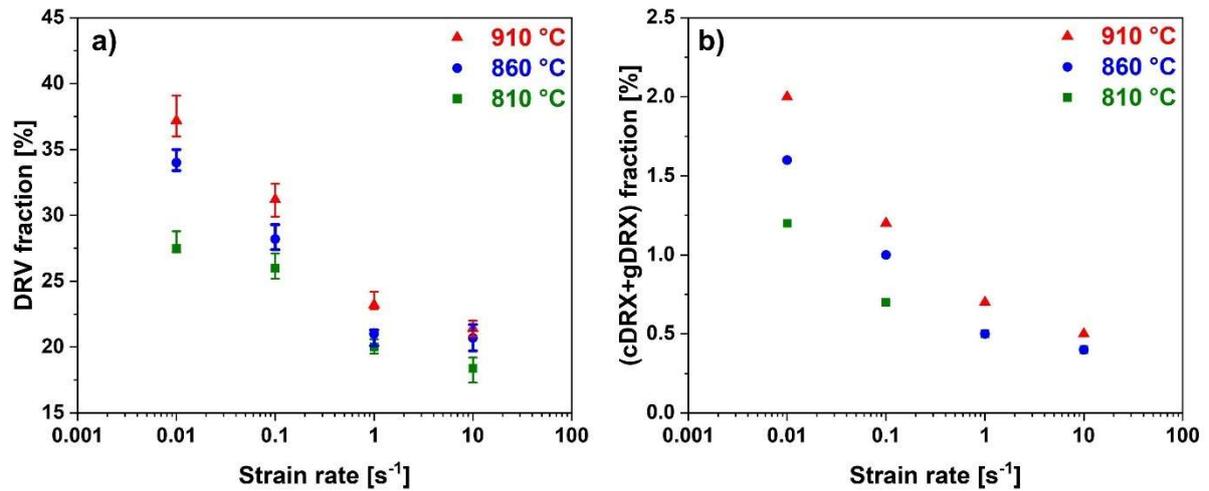

**Figure 7.** Area fraction of a) dynamic recovery, b) dynamic recrystallisation at different temperatures and strain rates (final strain = 0.85).

### 4.1.2 Subgrain size

The highlighted regions of the restored area, determined using the method introduced in 4.1.1, are converted into a partition to determine the average subgrain size. Since the fraction of the dynamically recrystallised area is negligible, we considered the restored area to be recovered for this purpose. Once the partition is created, the subgrain size distribution is plotted using the OIM software. Figure 8a shows the subgrain size distribution for compressions up to a final strain of 0.85 at 810 °C and various strain rates. In Figure 8a, the peaks and the fronts of distributions shift towards higher values when the strain rate decreases, indicating that lower strain rates promote the formation of larger subgrains. Similar behaviour was also found for the other studied temperatures. Figure 8b shows that increasing deformation temperature leads to a larger average subgrain size for a strain rate. The subgrain size distribution at other temperatures and strain rates are available in the supplementary data (Figures S4-S5).

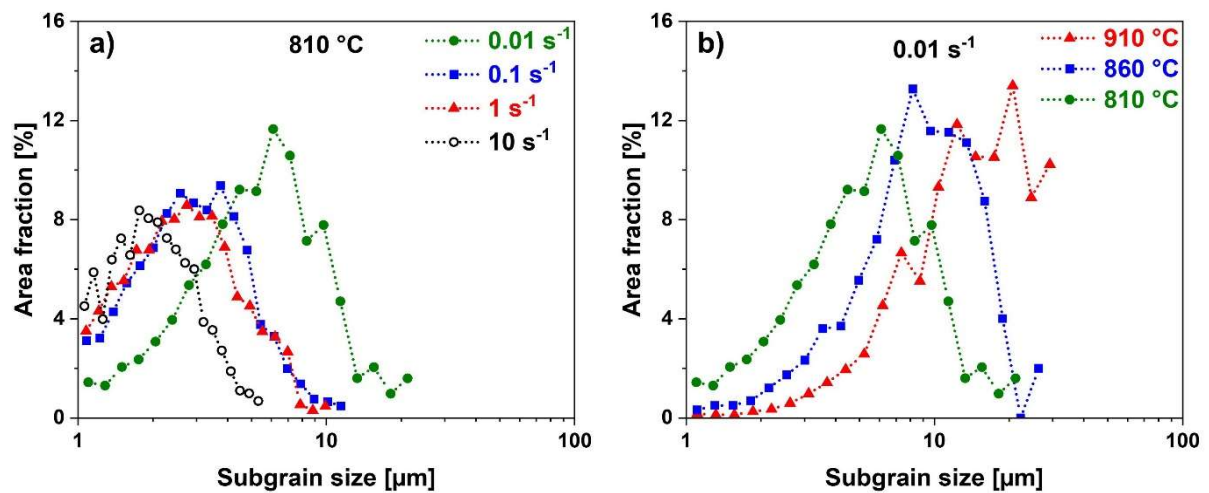

**Figure 8.** Subgrain size distributions corresponding to compressions up to a final strain of 0.85 at a) 810 °C and various strain rates, b) 0.01 s$^{-1}$ and different temperatures.

An alternative approach estimates the average subgrain size ($SG_S$) based on grain boundary density ($S_V$), as described by Eq. (3) [26].

$$SG_S = 2/S_V \quad (3)$$

The boundary density is calculated from the partition of the restored area using Eq. (4).

$$\begin{aligned} S_V &= (L_{LAGB} + L_{HAGB})/A_{DRV} \quad &\text{(a)} \\ A_{DRV} &= f_{DRV} \cdot A \quad &\text{(b)} \end{aligned} \quad (4)$$

where $L_{LAGB}$ and $L_{HAGB}$ represent the lengths of LAGBs and HAGBs, respectively. These lengths are determined from the restored fraction determined by the described process in the OIM software. $A_{DRV}$ and $f_{DRV}$ are the area and the fraction of the dynamically recovered area, respectively. Finally, $A$ is the total area measured using EBSD. Figure 9a shows the average subgrain size across different deformation conditions, as calculated using Eq. (3). Increasing the strain rate from 0.01 s$^{-1}$ to 0.1 s$^{-1}$ significantly decreases subgrain size. At strain rates higher than 0.1 s$^{-1}$, the average subgrain size exhibits only slight variation. Figure 9b also shows the average subgrain size at all tested conditions, obtained from the size distribution (Figure 8). The average subgrain size calculated using Eq. (3) (Figure 9a) suggests values and trends that are generally consistent with those obtained from subgrain size distribution (Figure 9b), validating the feasibility of using Eq. (3) for this analysis. As shown in Figure 9b, the average subgrain size varies from 2.2 µm at 810 °C and 10 s$^{-1}$ to 15.6 µm at 910 °C and 0.01 s$^{-1}$. The fine subgrains observed, particularly at higher strain rates, result from retarded dynamic recovery, possibly due to slow diffusion of Mo (see section 4.2 for further analysis).

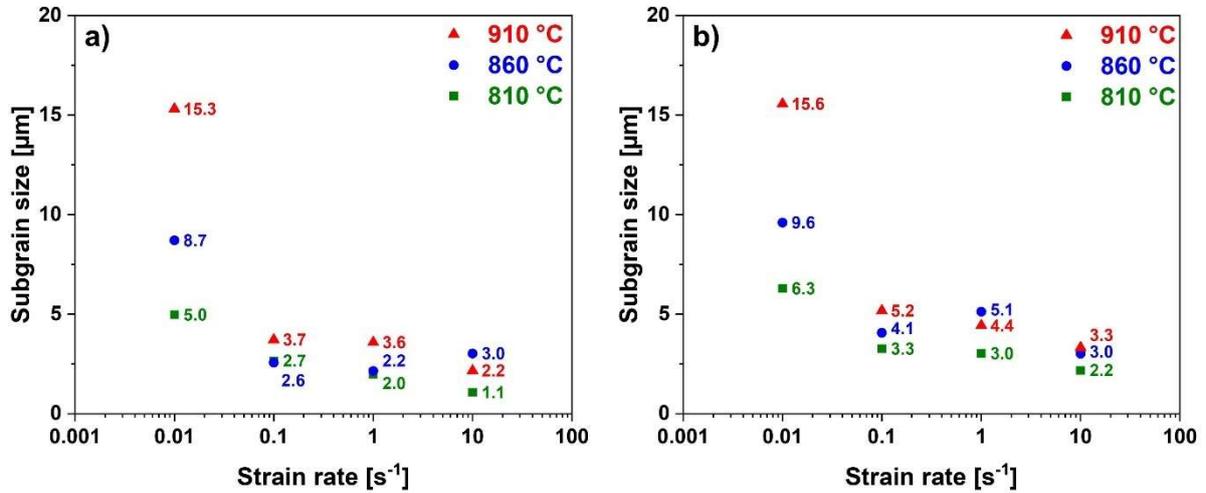

Figure 9. Average subgrain size corresponding to compressions up to a final strain of 0.85 at different temperatures and strain rates determined based on a) boundary density, b) size distribution.

### 4.1.3 The subgrain formation front

The quantification of DRV reveals that the subgrain formation in this material occurs as a necklace, similar to dDRX in low SFE materials. Due to the higher dislocation density near the prior grain boundaries, the subgrains are formed preferentially at these locations, particularly at higher strain rates. Before the steady-state flow stress is reached, further deformation does not change the average size of the recovered subgrains. However, subgrains continue to form and extend towards the grain interiors. Figure 10 schematically illustrates the evolution of subgrain formation.

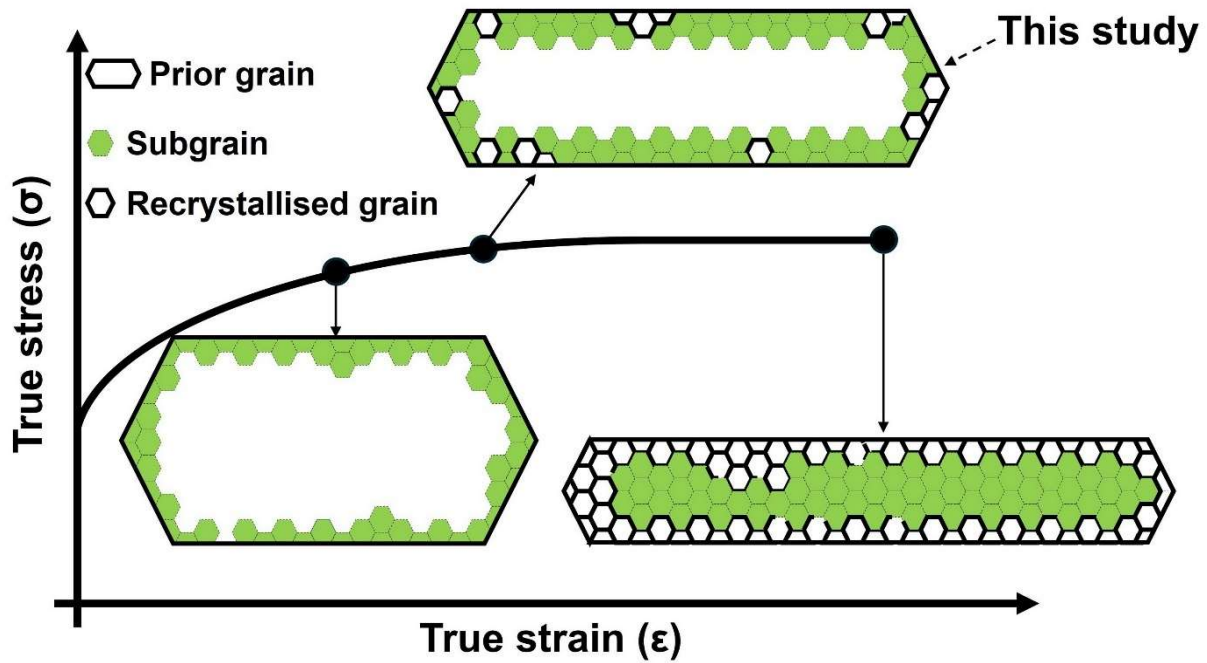

Figure 10. Schematic evolution of subgrain formation.

### 4.2 Influence of alloying elements

The observed broad work-hardening in flow stresses at strain rates exceeding 0.01 s$^{-1}$ is uncommon in most β-Ti alloys and can be attributed to the retardation of dynamic restoration mechanisms due to the influence of Mo atoms on the diffusion-controlled processes. Mo atoms retard the movement of dislocations since it is a slow-diffusing element [60] and pin the dislocations due to different atomic radii, eventually causing a delay in the rearrangement of dislocations into subgrains. Therefore, the dislocation pile-up and subgrain formation are more pronounced close to grain boundaries. The effect of Mo is more evident when the flow stress of Ti-15Mo is compared to that of Ti-17 (Mo equivalency of ~ 5%) at the same deformation condition. Figure 11a compares the flow stresses of the studied Ti-15Mo alloy and a Ti-17 [26] alloy after compression to 0.85 at 930 °C and 0.1 s$^{-1}$. Figure 11a shows that the flow stress of Ti-17 reaches a steady-state regime immediately after a slight discontinuous yielding. In contrast, in the flow curve of Ti-15Mo, no yielding occurs, and a broad work hardening is observed up to a plastic strain of 0.2. Figure 11b shows the evolution of the strain hardening with the strain for the two materials. The strain hardening is more pronounced in Ti-15Mo compared to Ti-17, reflecting fewer microstructural restoration. Furthermore, we compared the local misorientations within the Ti-15Mo and Ti-17 alloy microstructures from the EBSD data using kernel average misorientation (KAM) maps. Figure 12 displays KAM maps corresponding to samples of Ti-15Mo and Ti-17 alloys compressed to 0.25 final strain under the abovementioned condition (930 °C, 0.1 s$^{-1}$). We employed the sixth nearest neighbour KAM map for the Ti-15Mo and the first nearest neighbour KAM map for the Ti-17 to minimise the effect of different step sizes used in the EBSD measurements (0.35 µm for Ti-15Mo and 2 µm for Ti-17). This method allows for direct comparison of the two alloys. Although Ti-15Mo has finer initial grains, leading to a lower strain hardening exponent [61], it exhibits more pronounced misorientations than the Ti-17 alloy. The average KAM value determined for the Ti-15Mo alloy is 0.94, compared to 0.47 for Ti-17. Higher KAM values for Ti-15Mo imply higher dislocation density and retarded DRV during deformation. Both observed strain hardening exponents and determined KAM values provide evidence for the influence of Mo atoms on the deceleration of softening mechanisms. The fact that Mo retards the restoration mechanism has also been reported in recent studies. Lee et al. [42] reported that the higher amount of Mo in Ti-xMo-Fe delays the DRX. Similarly, the work of Momeni et al. [34] concludes that the larger amount of Mo in Ti4733 decreases the recovery rate compared to Ti5553. Ebied et al. [50] also observed similar broad peak stresses at high strain rates (1 s$^{-1}$) in a Ti-17Mo alloy, although the correlation with the effect of Mo remained unexplored.

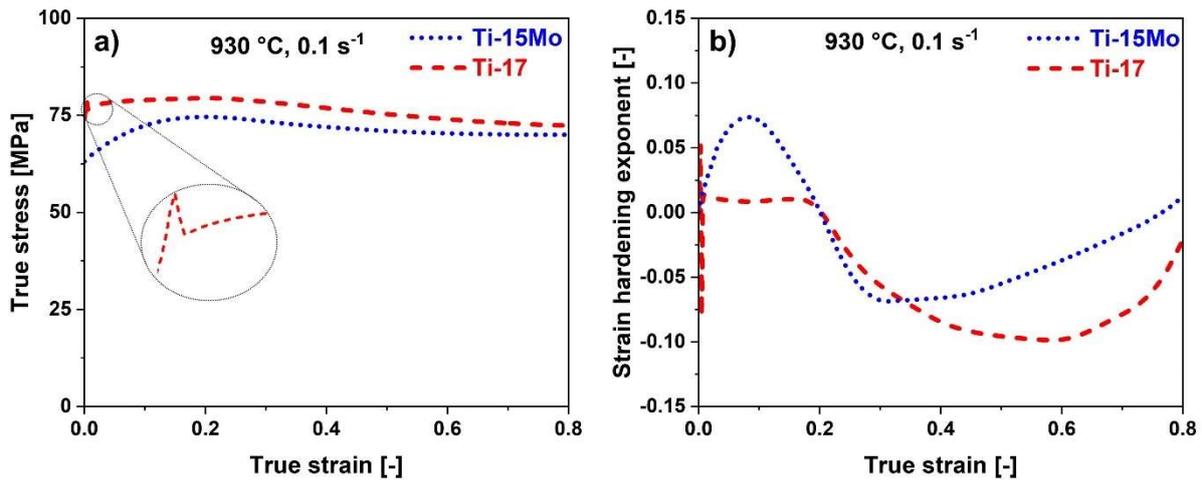

Figure 11. Flow behaviour in Ti-15Mo and Ti-17 alloys deformed at 930 °C and 0.1 s$^{-1}$, a) flow curve, b) strain hardening exponent.

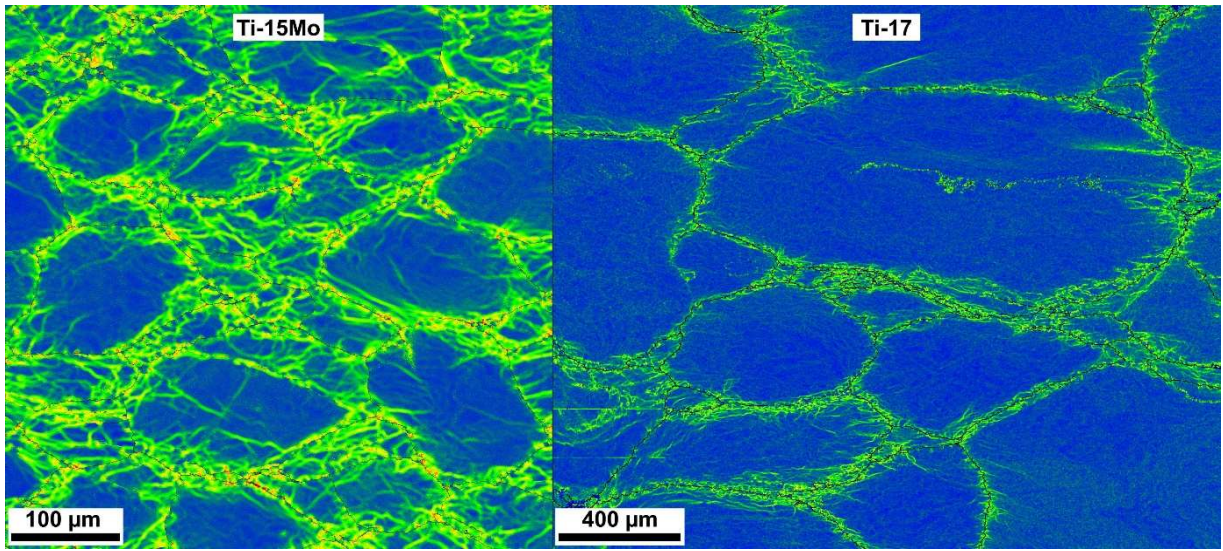

Figure 12. Kernel Average Misorientation (KAM) maps correspond to Ti-15Mo, and Ti-17 samples compressed at 930 °C and 0.1 s$^{-1}$ up to a final strain of 0.25. Compression axis is vertical.

## 5  Conclusions

This study employed hot compression experiments in the single β domain using a Gleeble 3800 simulator to investigate the flow behaviour and microstructural evolution of a Ti-15Mo alloy. Flow curves along with EBSD measurements, were used to uncover the primary restoration mechanisms. An EBSD-based method was developed to quantify dynamically recovered and recrystallised areas as well as subgrain size. The effect of Mo atoms on dynamic restoration mechanisms was investigated. The main conclusions are:

- The flow stress decreased at higher temperatures and lower strain rates. However, the relatively low values of the strain rate sensitivity slightly influence the strain rate.
- Strain hardening exponent calculations demonstrates the effect of strain rate on the retardation of softening mechanisms.
- EBSD measurements revealed a highly heterogeneous microstructure after deformation, where the DRV is the dominant mechanism across investigated temperatures and strain rates. However, new fine grains are formed by cDRX and gDRX. No evidence of dDRX was detected.
- The quantified DRV and DRX fractions decrease with the increasing strain rate and decreasing temperature. Notably, a very low fraction of DRX was observed, reaching up to 2% for the highest temperature and lowest strain rate.
- Two methods for measuring average subgrain size yielded consistent results, suggesting the presence of fine subgrains, possibly influenced by the low diffusivity of Mo atoms.
- A new concept of subgrain formation proposes the development of a necklace configuration of subgrains close to prior HAGBs in the early stages of deformation and extending into the grain interior at higher strains. Further deformation (e.g. torsion) is necessary to achieve a full subgrain structure.
- Slow diffusion of Mo and partial pinning of dislocations hinder softening mechanisms, as supported by the strain hardening exponent analysis and kernel average misorientation maps.

### Data availability

The raw and processed data are available for download from https://doi.org/10.3217/tzb6d-mfr03.

### Acknowledgements

This work was supported by the Austrian Science Fund (FWF) joint project, project no. I5818-N (deformation-phases-strength interaction in β-Ti alloys), and Czech Science Foundation (GACR), project no. 22-21151K. M. J., J. S., P. H., and D. P. gratefully acknowledge the partial financial support by OP Johannes Amos Copernicus of the MEYS of the CR, project No. CZ.02.01.01/00/22_008/0004591. We acknowledge Graz University of Technology for open access funding.